# An Interpretable Power System Transient Stability Assessment Method with Expert Guiding Neural-Regression-Tree

Hanxuan Wang, Na Lu*, Zixuan Wang, Jiacheng Liu, Jun Liu

Systems Engineering Institute, School of Automation Science and Engineering, Xi'an Jiaotong University, Xi'an, China 710049

## Abstract

Deep learning based transient stability assessment (TSA) has achieved great success, yet the lack of interpretability hinders its industrial application. Although a great number of studies have tried to explore the interpretability of network solutions, many problems still remain unsolved: (1) the difference between the widely accepted power system knowledge and the generated interpretive rules is large; (2) the probability characteristics of the neural network have not been fully considered during generating the interpretive rules; (3) the cost of the trade-off between accuracy and interpretability is too heavy to take. To address these issues, an interpretable power system Transient Stability Assessment method with Expert guiding Neural-Regression-Tree (TSA-ENRT) is proposed. TSA-ENRT utilizes an expert guiding nonlinear regression tree to approximate the neural network prediction and the neural network can be explained by the interpretive rules generated by the tree model. The nonlinearity of the expert guiding nonlinear regression tree is endowed with the extracted knowledge from a simple two-machine three-bus power system, which forms an expert knowledge base and thus the generated interpretive rules are more consistent with human cognition. Besides, the expert guiding tree model can build a bridge between the interpretive rules and the probability prediction of neural network in a regression way. By regularizing the neural network with the average decision length of ENRT, the association of the neural network and tree model is constructed in the model training level which provides a better trade-off between accuracy and interpretability. Extensive experiments indicate the interpretive rules generated by the proposed TSA-ENRT are highly consistent with the neural network prediction and more agreed with human expert cognition.

**Keywords:** Deep learning, Expert knowledge, Interpretability, Neural-regression-tree, Transient stability assessment

# 1 Introduction

Transient stability assessment is one of the most important topics in the research of operation, planning and control of power system [1-4]. Power system transient stability refers to the ability of power system to keep stabile after a set of contingencies. According to the estimation result of TSA, different operations can be executed to reduce the damage of these contingencies to the power system stability. However, in recent years, with the rapid development of renewable energy generation, smart grid and UHV transmission technology, the dynamic characteristic of power system has presented more variety [5]. The behaviors of power system become more and more unpredictable. With all these complications, TSA is becoming more difficult and time consuming gradually. In fact, when power system is under disturbance, it is necessary to assess the state of the transient stability as quickly as possible, so as to save enough time for emergency control.

Many methods have been developed for TSA including model-driven methods and data-driven methods mainly. Model-driven methods were usually constructed on the basis of time domain simulation or Lyapunov stability theory [6-8]. However, these methods might suffer from high computational cost or weak adaptability. Data-driven methods, such as SVM, do not need to model a specific problem and thus are faster and more adaptive [9]. Unfortunately, the data-driven methods highly rely on the engineered feature quality of the data and are influenced by the curse of dimension. With the rapid development of deep learning (DL) [10-12], DL based TSA methods get widely used without dependence on handcrafted features. A great number of DL based TSA methods have emerged [1, 2, 13-16]. Despite of the magnificent performance of deep learning in TSA, the decision logic of a neural network model is not transparent which makes the predictions uninterpretable. Due to the unpredictable risk, the lack of interpretability greatly hinders the practical application in industry.

To develop interpretable deep learning methods for TSA, a number of techniques have been applied to explain the neural network behaviors, including attention mechanism [17], decision tree [18, 19], Shapley [20] and so on. Among these techniques, decision tree (DT) based methods have caught extensive attention since it can generate decision logic directly. The main idea of DT based methods is to use the decision tree to mimic the prediction of the neural network and generate corresponding decision logic to explain the neural network behavior, which is called Neural Tree. Along this direction, many important studies have been performed. In Ref. [21], Wu et al. proposed a regional tree regularization which encouraged a deep model to be well-approximated by several separate decision trees specific to predefined regions of the input space. Song et al. proposed a distillation method called tree-like decision distillation to teach a student model the same problem-solving mechanism with the teacher model [22]. Specifically for TSA, Oliveira et al. proposed to assess power system operation security under multiple contingencies using a multiway decision tree for interpretability [19]. Ren et al. proposed an interpretable DL-based TSA model to balance the TSA accuracy and transparency by regularizing DL-based model with the average decision tree path length in the training process [18]. Although these methods can provide some interpretive rules for DL based TSA models, there are still some problems remaining unsolved: (1) in previous studies, the difference between the interpretive rules generated by linear decision trees and human cognition is quite large. (2) The decision tree models previously used mainly focus on the hard labels (stable or unstable) which ignores the probability information of the neural network. This ignorance results in information loss and incomplete interpretability. (3) Model performance gap between neural network and decision tree is huge. To generate effective interpretive rules, the neural network has to greatly sacrifice the accuracy.

To address the above problems, an interpretable power system transient stability assessment method with expert guiding neural-regression-tree (TSA-ENRT) is developed, which combines four modules including expert knowledge base, neural network evaluation model (NNEM), nonlinear regression tree (NRT) and tree regularization surrogate model. Firstly, expert knowledge is extracted from the algebraic equations in power flow calculation of a two-machine three-bus system in analogy to complex real system, and an expert knowledge base is constructed. Nonlinear terms of the original TSA features with clear physical meanings can be extracted from the expert knowledge base. These nonlinear terms together with the original TSA features make up the input to the tree model. Then, a neural network evaluation model can be trained with the original TSA features. Different from traditional TSA binary hard classification, we train the neural network evaluation model as a regression task to retain the probability information. By approximating the prediction of the evaluation model, an expert guiding nonlinear regression tree model can be constructed to mimic the probability generation behavior of the neural network. However, there is still a large gap between the performance of the neural network and the nonlinear tree model, even though the nonlinear tree model has stronger fitting ability compared with linear tree models. In order to further improve the consistency between the neural network evaluation model and the expert guiding nonlinear regression tree, the average tree depth is utilized to regularize the training of the neural network. A surrogate model is introduced to approximate the average tree depth which constructs a differentiable relation between the parameters of the neural network and the average tree depth. The combination of neural network, regression tree and average tree depth regularization is called neural-regression-tree. During the prediction process, the transient stability status of the power system and the corresponding probability can be reported by the well-trained neural network evaluation model and the decision logic can be generated from the expert guiding nonlinear regression tree. The contributions of this paper are summarized as follows:

1) Expert knowledge is introduced into the interpretability model. An expert knowledge base is built by extracting expert knowledge from a simple equivalent system. A nonlinear regression tree can be generated with the guide of the expert knowledge base. The decision logic generated by the nonlinear tree model with nonlinear terms of clear physical meaning can better simulate the nonlinear characteristics of the neural network.

2) The nonlinear regression tree can alleviate the cost of the trade-off between the accuracy and interpretability of the neural network model compared with previous neural tree solutions, because the nonlinear tree model can better approximate the nonlinear behavior of the neural network.

3) The TSA binary classification problem is transformed into a regression task of stability

probability in TSA-ENRT. Different decision logics can be generated by the nonlinear regression tree according to the probability predicted by the network model. The probability generation behavior of the neural network can be explained by the generated decision logic.

The remainder of this paper is organized as follows: Section II describes the problem formulation. Then the framework of the proposed TSA-ENRT is introduced in detail in Section III. Case studies and analysis are presented in Section IV. Section V gives the conclusions.

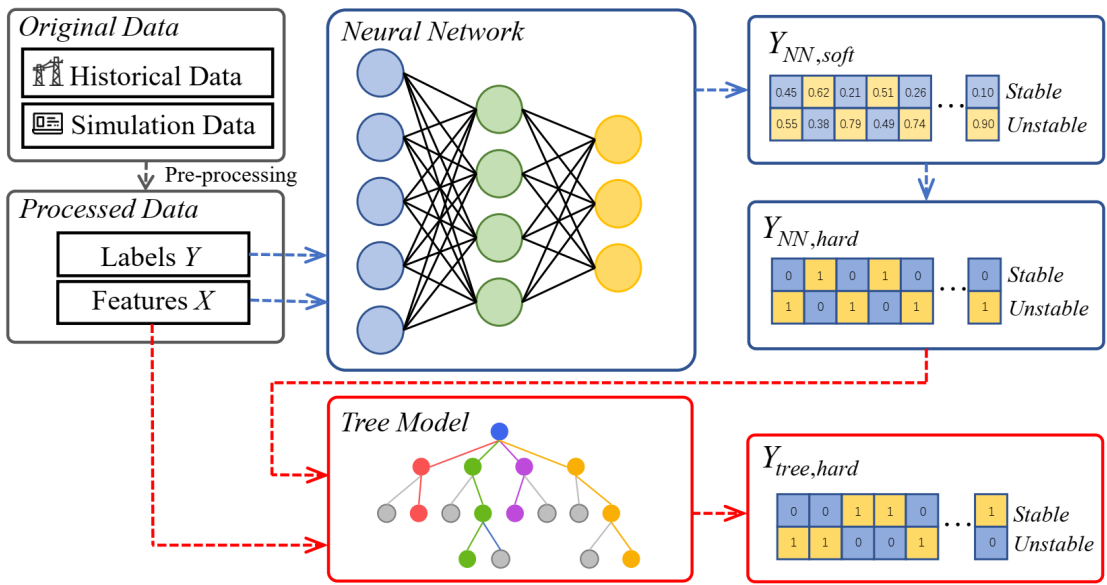

Fig. 1 The schema of the neural tree model.

## 2 Problem Analyses

Deep learning methods have shown remarkable performance in the area of transient stability assessment for their nonlinear feature extraction ability. To promote the industrial deployment of DL based TSA, combining interpretive tree model with neural network is a promising way, called neural tree. The paradigm of neural tree is shown in Fig. 1. A tree model, like decision tree, is built by approximating the prediction of the neural network. If the predictions of the neural network and the tree model are the same, i.e., $Y_{NN,hard,i}=Y_{tree,hard,i}$ for the $i^{th}$ sample as shown in Fig. 1, then the decision path of the tree model can be treated as the interpretive rule of the neural network decision logic. Based on this interpreting paradigm, the consistency between the tree model and the neural network vitally affects the interpretability of the neural tree. With higher consistency, more confident interpretive rules can be generated and the interpretability of the neural tree will be stronger. However, in previous neural tree studies [18, 19, 22, 23], the deployed tree models are mainly linear decision tree, and the difference of fitting capability between the neural network and the linear decision tree is very large. The fitting capability gap leads to low consistency and the generated interpretive rules become untrustworthy. Besides, it is also unreasonable to use the generated linear interpretive rules to explain a strongly nonlinear problem, i.e., transient stability assessment. Although some nonlinear tree models have been developed, their generated rules may include nonlinear terms without clear physical meanings and cannot benefit from human acknowledgment.

To further clarify our viewpoint, an exemplified neural tree has been built with a GRU neural network and a linear decision tree. The neural tree was trained and tested on CEPRI-TAS power system, and the detailed information of the system can be found in Section 4. The GRU model was trained for 60 epochs, and the linear decision tree is generated based on the prediction of the GRU model during each epoch. To quantitatively evaluate the consistency between the GRU model and the decision tree, fidelity $\rho$ has been introduced which can be expressed as [19, 23]

$$\rho = \frac{1}{N}\sum_{i=1}^{N} 1_{Y_{nn,hard,i}=Y_{tree,hard,i}}, \quad (1)$$

where $1_{Y_{nn,hard,i}=Y_{tree,hard,i}}$ is an indicator function, and when the hard label predicted by the neural network of the $i^{th}$ sample is the same as the tree model, $1_{Y_{nn,hard,i}=Y_{tree,hard,i}}$ equals to 1, otherwise 0. According to our experiment results, after 35 epochs of training, the accuracy of the neural network and the decision tree no longer changes, arriving at 97.3% and 75.0% respectively. The corresponding fidelity $\rho$ is 76.1%. Similar results have also been reported in [18] where the experiments were conducted on New England 10-machine 39-bus system. The accuracy of the neural network and the decision tree were 95% and 90% respectively, and the corresponding fidelity was about 87%. These results suggest that to get reliable interpretive rules from the tree model, the training of the neural network has to be terminated prematurely, and the learned decision boundary is very simple for the decision tree to approximate. Although the early stopping can improve the consistency, the accuracy of the neural network is relatively low at this moment. In other words, we have to sacrifice a great deal of neural network performance to improve the consistency which is unacceptable apparently. What's more, with the training process going on, the stability of some samples no longer changes, and as a result, the decision path generated by the tree model remains unchanged. However, the stable probability predicted by the neural network of these samples keeps changing. For example, the stable probability of a sample in the training set might has changed from 53% to 97%, but the corresponding interpretive rules haven't changed accordingly. Apparently, such interpretive rules fail to fully consider the probability information of the neural network. The reason behind this is that the inputs of the decision tree are supposed to be hard labels (stable or unstable). Therefore, the predicted probability of the neural network has to be rounded into one-hot label and the probability information gets discarded. In conclusion, we are facing three severe problems:

(1) lack of interpretive rules with clear physical meaning to explain the nonlinear behavior of the neural network;

(2) the interpretive rules cannot build connection with the neural network probability information;

(3) the trade-off between the neural network accuracy and the consistency (fidelity) between the neural network and the tree model is too large.

## 3 The Proposed Method

In this section, we first describe the overall framework of expert guiding neural-regression-tree in section 3.1. In section 3.2 and 3.3, the methods of constructing the expert guiding nonlinear regression tree and training the tree regularization surrogate model are described, respectively.

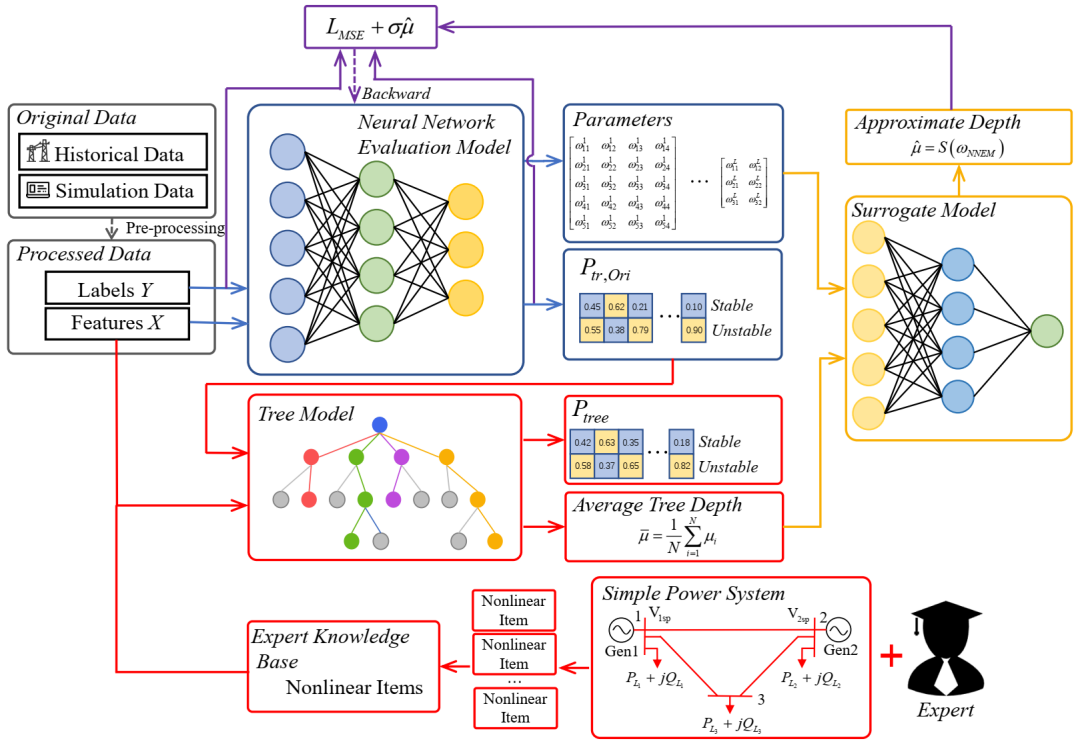

Fig. 2. The schema of TSA-ENRT.

### 3.1 Framework of Expert Guiding Neural-Regression-Tree

To address the above three issues, an interpretable power system transient stability assessment method with expert guiding neural-regression-tree (TSA-ENRT) is proposed as illustrated in Fig. 2.

TSA-ENRT consists of four parts: neural network evaluation model, nonlinear regression tree, expert knowledge base and tree regularization surrogate model. Since TSA is a binary classification problem, the original training set with $N$ samples can be denoted as $X_{tr,Ori} = \{(x_{tr,Ori,i}, y_{tr,Ori,i})\}_{i=1,\ldots,N}$. The original testing set $X_{te,Ori}$ can be defined in the same way. The nonlinear terms with clear physical meaning in power system can be abstracted from the algebraic equations in power flow calculation of a two-machine three-bus system which can compose an expert knowledge base of nonlinear features. Together with the original features, a nonlinear tree model can be generated based on the predicted stability status of the neural network evaluation model. As a result, the generated nonlinear tree model can approximate the nonlinear behavior of the neural network better compared with linear tree models and has better interpretability since the nonlinear interactions have clear physical meaning. The better fitting capability of the nonlinear tree model can alleviate the accuracy sacrifice for the consistency. The input and output of the evaluation model are the original TSA features and the stable probability respectively. If the nonlinear tree model is generated based on hard labels as the previous studies did [18, 19, 22, 23], the probability information will be ignored. Therefore, the generated interpretive rules cannot explain the inner probability inferring behavior of the neural network evaluation model. For this reason, we recast TSA as a regression task for both the neural network evaluation model and the nonlinear tree model. The objective function of the neural network evaluation model can be formulated as

$$L = \sum_{i=1}^{N} \left\| f_{NNEM}(x_{tr,Ori,i}) - y_{tr,Ori,i} \right\|_2^2 + \sigma_{tree} L_{tree}, \qquad (2)$$

where $f_{NNEM}$ is the neural network evaluation model. In order to establish the relation between the neural network evaluation model and the tree model, the average depth of the nonlinear regression tree is used for regularizing, called tree regularization, i.e., $L_{tree}$ in Eq. (2), and $\sigma_{tree}$ is a strength parameter for the tree regularization. However, the average tree depth is a undifferentiable scalar, and the parameters of the neural network evaluation model cannot benefit from the average tree depth. Inspired by [23], a surrogate model is introduced to build a mapping from the parameters of the neural network evaluation model to the average tree depth. Based on the surrogate model, the tree regularization becomes differentiable. We choose Multilayer Perception (MLP) as the surrogate model and the objective function of TSA-ENRT can be rewritten as

$$L_{NNEM} = \sum_{i=1}^{N} \left\| f_{NNEM}\left(x_{tr,Ori,i}\right) - y_{tr,Ori,i} \right\|_2^2 + \sigma_{tree} S\left(\omega_{NNEM}\right), \tag{3}$$

where $S(\cdot)$ is the surrogate model. The input of $S(\cdot)$ is the parameters of the neural network evaluation model and the output is the approximated average tree depth. Besides, we have also proposed a training strategy to prevent the surrogate model from overfitting, and the details can be referred in Section 3.3.

---

**Algorithm 1: Training Algorithm of TSA-ENRT.**

**Input:** Original Training set $X_{tr,Ori}$, Maximum training epoch $N_{max}$, tree regularization strength $\sigma_{tree}$.
**Output:** Well-trained neural network evaluation model, nonlinear regression tree.
1: Generate expert knowledge base and nonlinear interactions $X_{tr,Exp}$.
2: **for** $iter \in \{1,2,\cdots,N_{max}\}$ **do**
3: Train surrogate model $S(\cdot)$.
4: Train neural network evaluation model with Eq. (2) and the well-trained $S(\cdot)$.
6: Predict the stability probability $p_{tr,Ori}$ of the training set.
7: Generate nonlinear regression tree with $\{[X_{tr,Ori},X_{tr,Exp}],p_{tr,Ori}\}$.
8: **end for**
9: $p_{te,Ori}=f_{NNEM}(x_{te,Ori})$.
10: Generate interpretive rules by the using nonlinear regression tree with $[x_{te,Ori},x_{te,Exp}]$.

---

After successfully training, the well-trained neural network evaluation model and the corresponding expert guiding nonlinear regression tree can be deployed, the stability probability can be predicted by the neural network evaluation model and the corresponding decision logic can be generated by the nonlinear tree model by analyzing the consistency. The detailed steps of the training process of TSA-ENRT are summarized in Algorithm 1.

**3.2 Expert Guiding Nonlinear Regression Tree**

In order to generate the interpretive rules of the neural network evaluation model, a nonlinear regression tree is used for rule extracting in TSA-ENRT. In previous neural tree studies [18, 19, 22, 23], linear decision tree models were used to explain the behavior of the neural network, and the generated interpretive rules are linear. However, it is unreasonable to use linear rules to explain a nonlinear decision process, since too much information will be discarded [24]. Although some decision tree methods can be endowed with nonlinear property with nonlinear activation functions, the interpretability of these methods is still very limited and lacks clear physical meanings [25, 26]. For example, if the nonlinear activation is sigmoid, then a nonlinear item $e^{\delta_G}$ might appear in the interpretive rules, where $\delta_G$ is the generator power angle, but $e^{\delta_G}$ doesn't have a clear physical meaning for transient stability assessment. To solve this problem, we build an expert knowledge base to integrate domain knowledge correlated nonlinear characteristics into the tree model. By this means, the generated interpretive rules incorporate nonlinearity, which can better simulate the neural network evaluation model compared with linear trees. What's more, the nonlinear terms are inherently interpretable with expert knowledge.

Generally speaking, the expert knowledge base is supposed to be established based on real power systems. However, a real power system is too complex to deduce the explicit expression.

Luckily, for an expert with transient stability assessment knowledge, it is quite simple to formulate the algebraic equations in power flow calculation of a simple two-machine three-bus system as shown in Fig. 3. It is worth mentioning that the simple system and real power systems have the same basic electrical relations and are both networked systems. As a result, they share essential basic principles. Based on this, the knowledge extracted from the two-machine three-bus system can be extended to real complex situations. Due to page limit, the full explicit formulations are not given here, the detailed information can be referenced in [27]. The equations with nonlinear interactions are as following:

(1) *Transformation from d-q coordinate to natural coordinate*

$$\begin{cases} i_{G,d} = \sin\delta_G \cdot i_{G,x} + \cos\delta_G \cdot i_{G,y} \\ i_{G,q} = -\cos\delta_G \cdot i_{G,x} + \sin\delta_G \cdot i_{G,y} \end{cases}, \quad (4)$$

$$\begin{cases} v_{G,d} = \sin\delta_G \cdot v_{G,x} + \cos\delta_G \cdot v_{G,y} \\ v_{G,q} = -\cos\delta_G \cdot v_{G,x} + \sin\delta_G \cdot v_{G,y} \end{cases}, \quad (5)$$

where $i_{G,d}$, $i_{G,q}$, $v_{G,d}$, $v_{G,q}$ are the current and voltage in the *d-q* coordinate system for the two generators (Gen1 and Gen2), $\delta_G$ is the power angle.

(2) *Load power equation*

$$P_L + jQ_L = I_L^* V_L, \quad (6)$$

where $P_L$, $Q_L$ are the bus active load and reactive load, and $I_L^*$ and $V_L$ are the bus positive sequence current and voltage.

(3) *Load characteristic equation*

$$\begin{aligned} P_L + jQ_L &= P_L \cdot K_P + jQ_L \cdot K_Q \\ s.t.\ K_P &= \left(a_P V_L^2 + b_P V_L + c_P\right), K_Q = \left(a_Q V_L^2 + b_Q V_L + c_Q\right). \end{aligned} \quad (7)$$

Eq. (7) is the general form of the load characteristic. From Eq. (4) to Eq. (7), several nonlinear terms can be extracted, which are shown in Table 1.

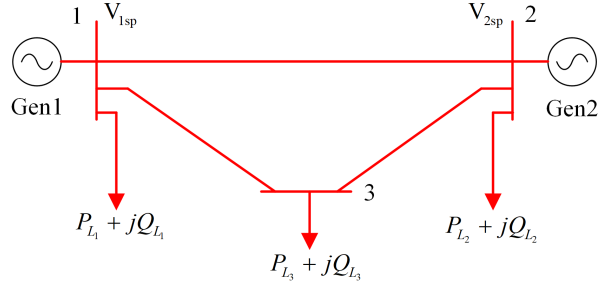

Fig. 3 The diagrams of the two-machine three-bus power system.

Table 1 Detailed information of the extracted nonlinear terms.

| Nonlinear Interaction | Physical Meaning |
|---|---|
| $\sin\delta_G$ | Sine transform of generator power angle |
| $\cos\delta_G$ | Cosine transform of generator power angle |
| $(P_B\cos\theta_B+Q_B\sin\theta_B)/V_B$ | Real part of bus positive sequence current |
| $(P_B\sin\theta_B-Q_B\cos\theta_B)/V_B$ | Imaginary part of bus positive sequence current |
| $V_B^2 P_B$ | Voltage quadratic transform of bus active load |
| $V_B^2 Q_B$ | Voltage quadratic transform of bus reactive load |
| $V_B P_B$ | Voltage linear transform of bus active load |
| $V_B Q_B$ | Voltage linear transform of bus reactive load |

Extending the nonlinear feature knowledge extracted from the two-machine three-bus system to real complex power systems and combining these nonlinear terms $X_{Exp}$ with the original features $X_{Ori}$, an extended dataset with nonlinear features $[X_{Ori}, X_{Exp}]$ can be generated for the nonlinear tree model. What's more, considering the problem (3) we declared in Problem Analyses, i.e. decision tree cannot simulate the probability behavior of the neural network evaluation model, we choose regression tree as the basis of the expert guiding nonlinear tree model to approximate the stability probability predicted by the neural network evaluation model. The mapping of the expert guiding nonlinear regression tree can be expressed as

$$P_{tree} = ERT\left(\left[X_{Ori}, X_{Exp}\right]\right), \tag{8}$$

where $P_{tree}$ is the stability probability predicted by the nonlinear tree model, $ERT$ is the well-trained expert guiding nonlinear regression tree. In this way, the generated interpretive rules can be integrated with expert knowledge to explain the predicted stability probability of the neural network evaluation model.

## 3.3 Tree Regularization

Despite we have built a nonlinear regression tree with expert knowledge, which can better approximate the neural network evaluation model compared with linear decision tree, the performance gap between the nonlinear tree and the neural network evaluation model is still large. There will be still an unacceptable sacrifice on the evaluation accuracy if we chase high consistency. To alleviate this trade-off dilemma, we utilize the average depth of the nonlinear regression tree to regularize the training of the neural network evaluation model, inspired by [18]. The main idea behind the tree regularization is simplifying the decision boundary of the neural network evaluation model. The simplified decision boundary is more friendly to the tree model and the performance of the tree model can be improved. The average depth $\bar{\mu}$ is the average value of decision path predicted by the tree model for all the training samples. The average depth can be obtained via

$$\bar{\mu} = \frac{1}{N} \sum_{i=1}^{N} \mu_i, \qquad (9)$$

where $\mu_i$ is the depth of the generated decision path corresponding to the $i^{th}$ sample and $\mu_i$ is defined as

$$\mu_i = PathLength\left(\left[x_{Ori,i}, x_{Exp,i}\right]\right). \qquad (10)$$

The function *PathLength* counts the number of nodes from the root node to the leaf node in the decision path. However, there is no explicit relationship between the parameters of the neural network evaluation model and the average depth $\bar{\mu}$. Apparently, $\bar{\mu}$ is undifferentiable. For above considerations, a surrogate model is introduced to establish a mapping from the neural network evaluation model parameters to the average depth of the nonlinear regression tree. The surrogate model can be realized using multilayer perception (MLP), and the mapping can be

expressed as

$$\hat{\mu} = S(\omega_{NNEM}), \qquad (11)$$

where $\hat{\mu}$ is the approximate average tree depth, $\omega_{NNEM}$ is the parameters of the neural network evaluation model and $S(\cdot)$ is the surrogate model mapping. Since the MLP is differentiable, the tree regularization can be easily integrated into the training objective function. Flattening all the parameters $\omega_{NNEM}$ into a vector as the input of the surrogate model, and the training label is the average tree depth $\bar{\mu}$. The objective function for training the surrogate model is formulated as

$$L_{surr} = \frac{1}{N_{surr}} \sum_{k=1}^{N_{surr}} \|\bar{\mu}_k - S(\omega_{NNEM})\|_2^2 + \sigma_{Surr} \|\omega_{surr}\|_2^2, \qquad (12)$$

where $N_{surr}$ is the number of the surrogate model training samples, $\sigma_{surr}$ represents the regularization coefficient and $\omega_{surr}$ is the parameters of the surrogate model.

Since the surrogate model aims to realize a mapping from the neural network evaluation model parameters to the average tree depth, its performance can directly affect the effectiveness of the tree regularization and it is very important to guarantee that the surrogate model can approximate the average tree depth accurately. According to Algorithm 1, during the training of the neural network evaluation model, a nonlinear tree will be generated based on the prediction of the training set after each epoch. With Eqns. (9) and (10), the average tree depth $\bar{\mu}$ can be acquired. The neural network evaluation model parameters and the corresponding average tree depth in each epoch will be collected to construct the training set of the surrogate model. In other words, the number of the surrogate model training samples equals to the training epoch $K$. However, the input of the surrogate model is the parameters of the neural network evaluation model, which is relatively large dimensional input. Therefore, if the training data of the surrogate model is insufficient, it can lead to overfitting and cause inaccurate approximations of the average tree depth. This in turn can affect the consistency between the neural network evaluation model and the tree model. Unfortunately, in the previously studies [18, 19, 22, 23], this situation has not been fully considered. Another problem that needs be considered is reusing inaccurate training data generated in the early stages will impact the performance of the surrogate model. However, the training data generated in the early stages are not accurate since the neural network evaluation model hasn't been fully trained. According to [28, 29], the surrogate model will tend to memorize the inaccurate information provided earlier, and treat the later generated samples which are more accurate as noisy information. These two issues can lead to severe overfitting and the average tree depth can hardly be approximated accurately.

For the above two problems, we propose a surrogate model training policy which combines sample selecting and data augmentation strategy. Due to insufficient training samples for the surrogate model, a base sample size $B$ is introduced for training the surrogate model. If the number of the surrogate model training samples is smaller than $B$, stochastic noise is added to the parameters of the neural network evaluation model for data augmentation, which can be expressed as

$$\omega_{Aug} = \omega_{NNEM} + \omega_{noise}, \tag{13}$$

where $\omega_{noise}$ is the noise sampled from a Gaussian distribution $\kappa$ with mean $\lambda$ and variance 0.01. $\lambda$ can be defined with

$$\lambda = \frac{1}{Num_{NNEM}} \sum_{i=1}^{Num_{NNEM}} \omega_{NNEM,i}, \tag{14}$$

where $Num_{NNEM}$ is the number of the parameters in the neural network evaluation model. The corresponding average depth $\bar{\mu}_{Aug}$ can be calculated by Eqns.(9) and (10). By this means, the issue of insufficient sample size can be alleviated and we have shown the performance of the surrogate model in Section 4.3. For the repeated training issue, when the training epoch is larger than $B$, we choose to assign different weights to the samples from different training periods. In detail, suppose the current epoch is $K$, for the surrogate model training samples whose corresponding epoch is smaller than $K/2$, the weight in the objective function is $1/K$, and for samples whose epoch is larger than $K/2$, the weight is 1. The modified objective function is defined as

$$L_{surr} = \sigma_{Surr} \|\omega_{Surr}\|_2^2 + \frac{1}{(\lfloor K/2 \rfloor)} \frac{1}{K} \sum_{k=1}^{\lfloor K/2 \rfloor} \|\bar{\mu}_k - S(\omega_{NNEM,k})\|_2^2 + \frac{1}{(K - \lfloor K/2 \rfloor)} \sum_{k=\lfloor K/2 \rfloor}^{K} \|\bar{\mu}_k - S(\omega_{NNEM,k})\|_2^2, \tag{15}$$

The detailed steps of training the surrogate model are summarized in Algorithm 2.

---

**Algorithm 2: Training Algorithm of the Surrogate Model.**

**Input:** Current training epoch $K$, Surrogate model training samples $ST = \left\{ (\omega_{NNEM,k}, \bar{\mu}_k) \mid \omega_{NNEM,k} \in R^{Num_{NNEM}}, \bar{\mu}_k \in R \right\}_{k=1}^{K}$, Stochastic noise $\kappa$, Regularization coefficient $\sigma_{surr}$, Base sample size $B$, Surrogate model $S(\cdot)$.
**Output:** Well-trained surrogate model $S(\cdot)$.
1: **if** $K < B$
2:  *Counter = K*;
3:  **while** *Counter < B*
4:    Generate a new sample $(\omega_{Aug}, \bar{\mu}_{Aug})$ by Eq. (9) and Eq. (10);
5:    $ST = ST \cup (\omega_{Aug}, \bar{\mu}_{Aug})$;
6:    *Counter=Counter*+1;
7:  **end while**
8:  Train surrogate model $S(\cdot)$ with Eq. (15) as the objective function until convergence;
9:  **return** $S(\cdot)$.
10:**else**
11: Train surrogate model $S(\cdot)$ with Eq. (15) as the objective function until convergence;
12: **return** $S(\cdot)$.
13:**end if**

---

## 4 Case Studies and Analyses

In this section, a regional power system in China is taken as an example to evaluate the performance of the proposed TSA-ENRT, where plenty of experiments have been conducted. Experiments on model performance, surrogate model training visualization, hyper-parameter sensitivity analysis, training process visualization, nonlinear interaction analysis and interpretive rule visualization have been performed.

## 4.1 Dataset and experiment setup

All experiments are conducted on a server with Intel Xeon CPU Gold 6230 of 2.1 GHz and TITAN V GPU. The proposed method is implemented in Python with deep learning framework Pytorch 1.12.1 and machine learning framework Scikit-Learn 1.1.

The test example is a simplified regional power system in China, named CEPRI-TAS, which is a 15-machine 85-bus system as shown in Fig 4. Monte-Carlo method has been used to generate samples with different initial state under specific contingency. The contingencies are three-phase faults with inter-area corridor trip and cleared within 150ms. We choose generator and bus features as the original TSA features, which are shown in Table 2. The data is simulated by BPA and the stability is discriminated like previous methods [1, 2, 15, 17, 18]. 2,000 samples were generated and the training set and testing set were split by the ratio of 3:1.

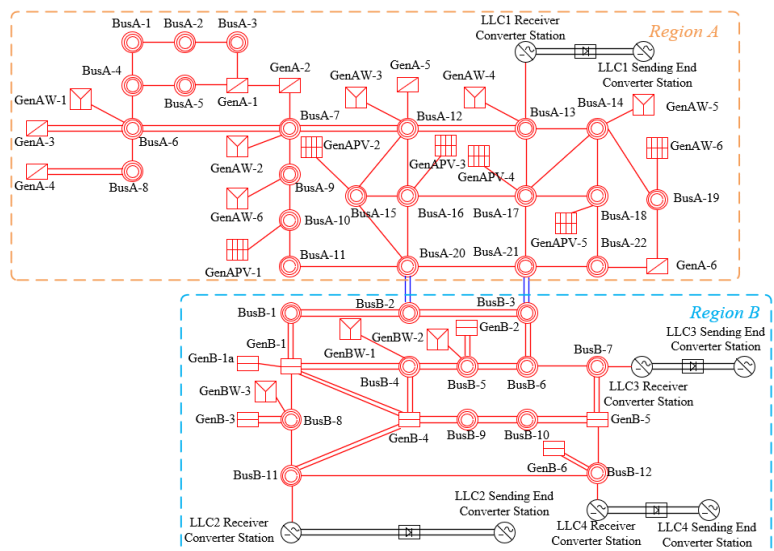

Fig. 4 The wiring diagrams of CEPRI-TAS power system.

For the experiment setup, we select GRU [30] as the neural network evaluation model, and the input size, hidden unit size and hidden layer of the GRU are 515, 200, 2 respectively, where there are 126,452 parameters. The structure of the MLP surrogate model is 126,452-1,000-25-1. During the training process, the base sample size $B$ for training the surrogate model is 20. Adam is used as the optimizer for both the neural network evaluation model and the surrogate model. The learning rates of the two models are both 0.001. The exponential decay rate for the first and second moment estimates $\beta_1$ and $\beta_2$ are set as 0.9 and 0.99 respectively.

Table 2 Detailed information of generator and bus features.

| Object | Number | TSA Feature (unit) | Symbol |
|---|---|---|---|
| Generator | 15 | Power Angle (degree) | $\delta_G$ |
| | | Speed Deviation (Hz) | $f_{G,SD}$ |
| | | Mechanical Power (MW) | $P_{G,MP}$ |
| | | Electromagnetic Power (MW) | $P_{G,EP}$ |
| | | Acceleration Power (MW) | $P_{G,AP}$ |
| | | Reactive Power (MVar) | $Q_{G,RP}$ |
| Bus | 85 | Positive Sequence Voltage (p.u.) | $V_{B,PSV}$ |
| | | Positive Sequence Angle (degree) | $\theta_B$ |
| | | Frequency Deviation (Hz) | $f_{B,FD}$ |
| | | Active Load (MW) | $P_{B,AL}$ |
| | | Reactive Load (MVar) | $Q_{B,RL}$ |

**4.2 Model Performance**

The comparison results on accuracy of TSA-ENRT and the related methods are reported in Table 3. The related methods include GRU-$L_1$, GRU-$L_2$ and GRU-TR [18]. Each method has two comparison candidates, for example, GRU-$L_1$ is the GRU model with $L_1$ regularization, GRU-$L_1$-RT is the corresponding regression tree which approximates the predictions of GRU-$L_1$. GRU-TR [18] is also an interpretable DL-based model with tree regularization in the training process. Although in this study we focus on the neural tree interpretability in the area of TSA, the accuracy is still one of the most important metrics. Compared with traditional methods (GRU-$L_1$ and GRU-$L_2$), the accuracy of TSA-ENRT has slightly decreased. This is due to the compromise in accuracy brought by the introduction of the tree regularization. For the comparison between the two tree regularization methods, the accuracy of the neural network evaluation model in TSA-ENRT is higher than GRU-TR across different levels of the tree regularization strength. The outperformance proves that our proposed method can effectively alleviate the accuracy and consistency trade-off and can achieve better performance in TSA tasks. For the tree model in each method, we noticed that the performance of the tree models with tree regularization based methods are superior to the traditional methods. It is worth mentioning that there is no comparability between the regularization strength of the traditional methods and tree regularization based methods. As a result, from Table 3, we can find that for $L_1$ and $L_2$ methods, when the regularization strength is set to $10^1$, the tree model with $L_2$ regularization achieves the highest accuracy, reaching up to 90.1% and the accuracy of the corresponding neural network is 87.8%. In TSA-ENRT and GRU-TR, when the tree regularization strength is $10^0$, the tree models achieve the highest accuracy 92.6% and 89.0% respectively, increasing by 2.5% and -1.1%. The accuracy of the corresponding neural networks is 6.5% and 5.3% higher than the $L_2$ method. These results demonstrate that TSA-ENRT can effectively balance the performance of the neural network evaluation model and the tree model.

Table 3 Performance of different methods on Accuracy.

| Method | Strength of Regularization | | | | | |
| --- | --- | --- | --- | --- | --- | --- |
| | $10^{-3}$ | $10^{-2}$ | $10^{-1}$ | $10^0$ | $10^1$ | $10^2$ |
| TSA-ENRT | 0.971 | 0.971 | 0.953 | 0.943 | 0.869 | 0.421 |
| TSA-ENRT-RT | 0.832 | 0.832 | 0.882 | 0.926 | 0.846 | 0.430 |
| GRU-TR [18] | 0.970 | 0.971 | 0.945 | 0.931 | 0.792 | 0.377 |
| GRU-TR-RT [18] | 0.793 | 0.794 | 0.817 | 0.890 | 0.805 | 0.370 |
| GRU-$L_2$ | 0.971 | 0.973 | 0.972 | 0.916 | 0.878 | 0.582 |
| GRU-$L_2$-RT | 0.750 | 0.750 | 0.750 | 0.767 | 0.901 | 0.593 |
| GRU-$L_1$ | 0.963 | 0.967 | 0.899 | 0.590 | 0.398 | 0.419 |
| GRU-$L_1$-RT | 0.750 | 0.749 | 0.759 | 0.595 | 0.383 | 0.410 |

Table 4 presents the performance of different methods on fidelity which is calculated with

$\frac{1}{N}\sum_{i=1}^{N} 1_{Y_{NNEM,hard,i}=Y_{ERT,hard,i}}$, where $Y_{NNEM,hard,i}$ and $Y_{ERT,hard,i}$ are the $i^{th}$ hard prediction of the neural network evaluation model and the nonlinear regression tree. It can be seen that the fidelity of the tree regularization based methods (TSA-ENRT, GRU-TR) is much higher than traditional methods. The outperformance indicates that the tree regularization can constraint the mapping relation between the neural network parameters and the average depth of the tree model. This constraint simplifies the decision boundary of the neural network evaluation model and make it more friendly to the tree model. Besides, the expert knowledge in the nonlinear regression tree endows TSA-ENRT with better fidelity than GRU-TR. This indicates that due to the existence of expert guiding nonlinear interactions, the shortcoming of the linear decision tree of insufficient to fit the nonlinear decision boundary has been made up and the trade-off on accuracy brought by the tree regularization could be alleviated. In other words, at the same level of consistency, TSA-ENRT needs the least accuracy to compromise. With the regularization strength increasing, fidelity of TSA-ENRT slowly increases from 0.847 to 0.991. This implies the decision boundary predicted by the neural network evaluation model becomes simpler gradually. However, when the tree regularization strength is very large, such as $10^2$, the accuracy of the neural network evaluation model becomes 0.421, which is even lower than random initialization, and the decision boundary is too simple to meet the TSA task. Therefore, excessively large tree regularization strength is beneath consideration.

Table 4 Performance of different methods on fidelity.

| Method | Strength of Regularization | | | | | |
|---|---|---|---|---|---|---|
|  | $10^{-3}$ | $10^{-2}$ | $10^{-1}$ | $10^{0}$ | $10^{1}$ | $10^{2}$ |
| TSA-ENRT | 0.847 | 0.847 | 0.913 | 0.971 | 0.974 | 0.991 |
| GRU-TR | 0.810 | 0.811 | 0.845 | 0.939 | 0.958 | 0.993 |
| GRU-$L_2$ | 0.762 | 0.761 | 0.765 | 0.831 | 0.957 | 0.989 |
| GRU-$L_1$ | 0.765 | 0.759 | 0.807 | 0.900 | 0.959 | 0.946 |

**4.3 Surrogate Model Training Visualization**

In the previous studies on neural tree [18, 22, 23], researchers mainly utilize neural network parameters and average tree depth which are generated during training process to train the surrogate model. Since the input dimension of the surrogate model equals the parameter size of the neural network evaluation model, the hyperparameters in the neural network evaluation model can directly influence the model size of the surrogate model. The corresponding relationship between the neural network evaluation model (GRU here) hyperparameters and the surrogate model input dimension has been shown in Fig. 5. It can be seen that, the parameter size of the surrogate model is enormous, and it is prone to overfitting when the training data is insufficient. In TSA-ENRT, data augmentation and reweight techniques have been employed to alleviate the overfitting dilemma. The data augmentation was implemented by adding noise to the parameters of the neural network evaluation model. The added noise is sampled from Gaussian distribution or Dirichlet distribution to expand training samples. If the overfitting can be alleviated, the surrogate model can predict the approximated average tree depth more accurately. Fig. 6 shows the relationship between the true average tree depth and the approximating tree depth predicted by the surrogate model from different training paradigms. The blue dotted line represents the true average tree depth $\bar{\mu}$ of the tree model during the training process while the blue solid line is the approximated average tree depth curve obtained by polynomial fitting based on $\bar{\mu}$. For convenience, the strength of the tree regularization is set as $10^{-2}$ to ensure the consistency of $\bar{\mu}$ among different training paradigms. Fig. 6(a) depicts the approximating average tree depth $\hat{\mu}$ predicted by the surrogate model using augmentation with Gaussian distribution noise, (b) is the result with augmentation of Dirichlet distribution noise, and the surrogate model in (c) is obtained following the previous methods [18, 22, 23]. It can be seen that the approximation performance of TSA-ENRT is much better than the previous methods. The average error of training surrogate model without data augmentation is 1.122, while the TSA-ENRT using augmentation with Gaussian and Dirichlet distribution are 0.722 and 0.726 respectively.

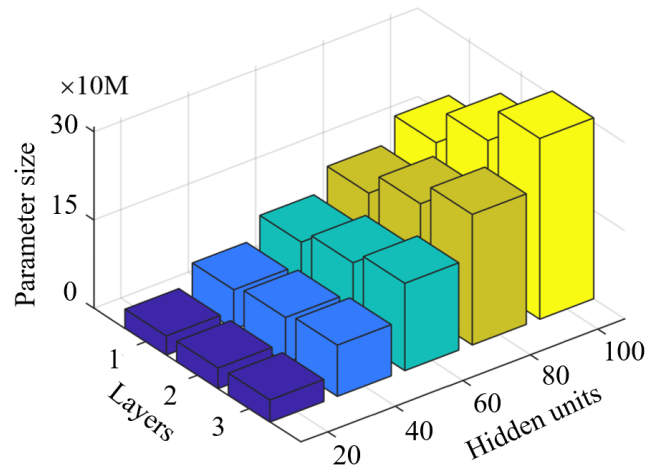

Fig. 5 The corresponding relation between the neural network evaluation mode hyperparameters and the surrogate model input dimension.

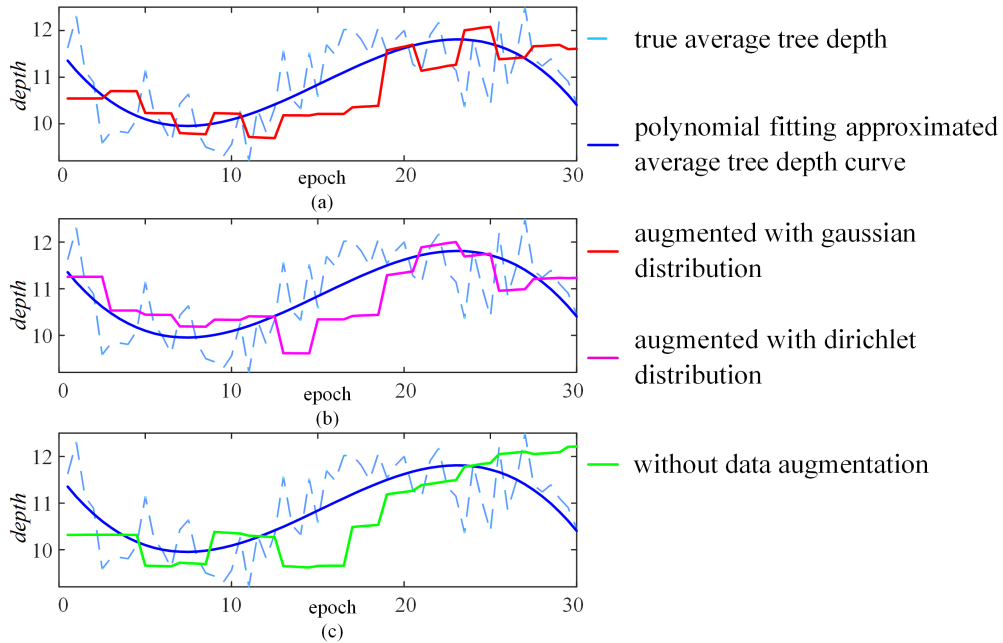

Fig. 6 The relationship between the true average tree depth $\bar{\mu}$ and the approximated tree depth $\hat{\mu}$ predicted by the surrogate model with different training paradigms, (a) augmentation with Gaussian distribution; (b) augmentation with Dirichlet distribution; (c) without data augmentation.

Besides, we have also analyzed the influence of the surrogate model training strategies on the performance of TSA-ENRT as shown in Table 5 and Table 6. The training strategies include reweight, data augmentation with Gaussian noise and Dirichlet noise. The strength of the tree regularization is set as $10^{-2}$. For a fair comparison, the contributions have been analyzed separately in scenarios with and without nonlinear interaction features. It can be seen that no matter there are nonlinear interactions or not, reweight and data augmentation can improve the performance. However, the contribution mainly reflects in the accuracy of the tree model and consistency rather than the neural network evaluation model (NNEM). For reweight, compared the results of the first and fourth rows in Tables 5 and 6, the effectiveness can be validated. For data augmentation, the comparisons among row 1, 2, 3 and 4, 5, 6 demonstrate that this strategy can improve the fidelity. What's more, the contributions of the Gaussian noise and the Dirichlet noise are similar. When both strategies are deployed in the training of the surrogate model, the tree model has the best performance, indicating the effectiveness of our designed strategies.

Table 5 Performance of TSA-ENRT with different surrogate model training strategies with nonlinear interactions.

| Reweight | Gaussian | Dirichlet | Acc | | Fidelity |
| --- | --- | --- | --- | --- | --- |
| | | | NNEM | Tree model | |
| × | × | × | 0.970 | 0.818 | 0.831 |
| × | × | ✓ | 0.970 | 0.822 | 0.836 |
| × | ✓ | × | 0.970 | 0.823 | 0.836 |
| ✓ | × | × | 0.971 | 0.827 | 0.838 |
| ✓ | × | ✓ | 0.971 | 0.832 | 0.847 |
| ✓ | ✓ | × | 0.970 | 0.831 | 0.847 |

Table 6 Performance of TSA-ENRT with different surrogate model training strategies without nonlinear interactions.

| Reweight | Gaussian | Dirichlet | Acc | | Fidelity |
| --- | --- | --- | --- | --- | --- |
| | | | NNEM | Tree model | |
| × | × | × | 0.970 | 0.793 | 0.810 |
| × | × | ✓ | 0.968 | 0.803 | 0.817 |
| × | ✓ | × | 0.969 | 0.811 | 0.822 |
| ✓ | × | × | 0.970 | 0.800 | 0.813 |
| ✓ | × | ✓ | 0.969 | 0.814 | 0.823 |
| ✓ | ✓ | × | 0.971 | 0.817 | 0.829 |

## 4.4 Hyper-parameter Sensitivity Analysis

In TSA-ENRT, the hyper-parameters include the strength of tree regularization $\sigma_{tree}$, the minimum number of samples required to be at a leaf node $S_{leaf}$ and the $L_2$ regularization strength $\sigma_{surr}$ for surrogate model training.

First, we have analyzed the influence of the $L_2$ regularization strength for the surrogate model training on accuracy and fidelity performance. The relationship between $\sigma_{surr}$ and the performance of TSA-ENRT is shown in Fig. 7. Fixing tree regularization strength and $S_{leaf}$ as $10^0$ and 10, when $\sigma_{surr}$ equals to $10^{-2}$ and $10^0$, there aren't significant changes on the performance of TSA-ENRT including accuracy and fidelity. However, when the strength of the surrogate model $L_2$ regularization is high, the accuracy of the neural network evaluation model increases but the accuracy of the tree model and fidelity decreases. We found that $\sigma_{surr} = 10^0$ leads to similar performance of TSA-ENRT with the tree regularization strength $\sigma_{tree}$ less than $10^{-2}$. The reason behind this is that too large $L_2$ regularization for the surrogate model will make the surrogate model degenerate. When the approximated average tree depth becomes so small ($\hat{\mu} < 0.2$) that the tree regularization can hardly make contribution to the optimization of TSA-ENRT. As a result, even if the tree regularization strength $\sigma_{tree}$ is relatively large, it can only contribute very small gradients to the training process.

$S_{leaf}$ and $\sigma_{tree}$ are directly related with the nonlinear regression tree model in TSA-ENRT, and it is necessary to analyze the influence of the two hyper-parameters on the performance. Since $\sigma_{tree}$ has already been studied in Section 4.3, fixing $\sigma_{tree} = 10^0$ and $\sigma_{surr} = 10^{-2}$, Fig. 8 shows the variation of TSA-ENRT performance on accuracy and fidelity with respect to $S_{leaf}$. It can be seen that too large $S_{leaf}$ will lead to the over-simplification of the nonlinear regression tree, making the mapping from the parameters of the neural network evaluation model to the average depth no longer accurate, decreasing the accuracy and fidelity of the proposed TSA-ENRT. The experiment results indicate that the selection of hyper-parameters should be conservative.

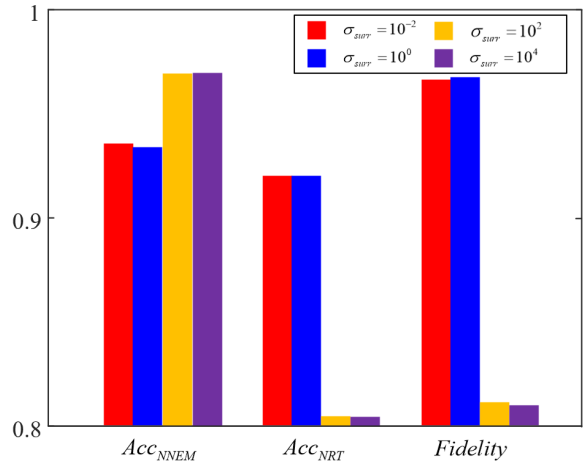

Fig. 7. The variation of fidelity with different surrogate model regularization strength.

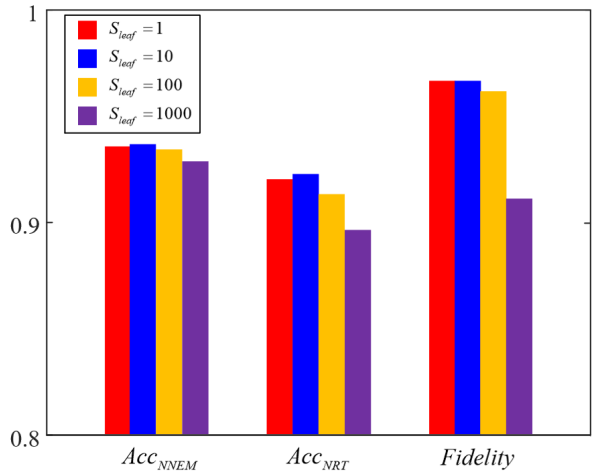

Fig. 8 The variation of TSA-ENRT performance on accuracy and fidelity with respect to $S_{leaf}$.

### 4.5 Nonlinear Term Analysis

The expert knowledge base is one of the most important parts of TSA-ENRT. Therefore, it is necessary to analyze the characteristics of the nonlinear terms. We have extracted 8 nonlinear terms from the algebraic equations for power flow calculation based on a simple two-machine three-bus system, as shown in Table 1. The nonlinear terms will be analyzed from three aspects:

(1) the frequency of nonlinear term occurrence in interpretive rules, (2) the amount of information of nonlinear terms in the nonlinear regression tree and (3) the influence of different nonlinear term combinations on the performance of TSA-ENRT. In these experiments, the strength parameter of tree regularization is $10^0$, and corresponding neural network evaluation model accuracy and fidelity are 0.943 and 0.971 respectively. Under this parameter setting, a good balance between accuracy and consistency can be achieved.

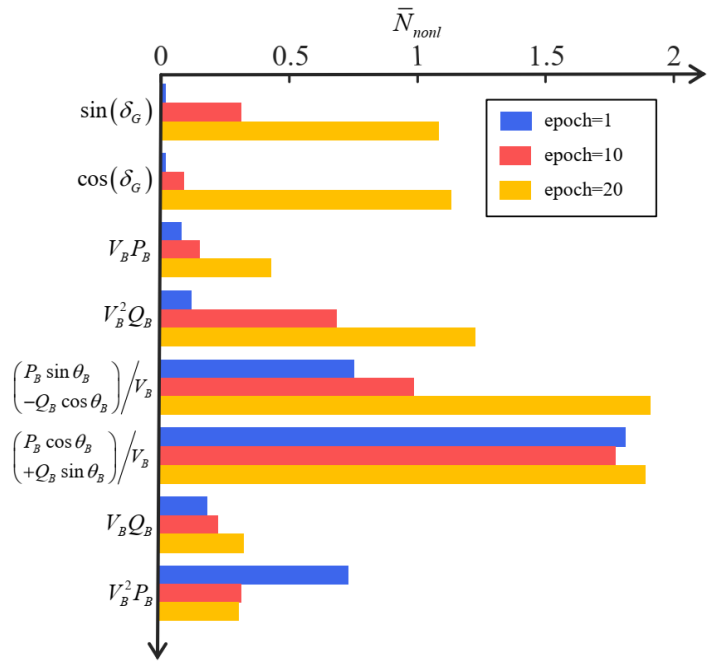

Fig. 9 The average frequency $\overline{N}_{nonl}$ of the nonlinear interactions in interpretive rules.

Firstly, Fig. 9 shows the average frequency $\overline{N}_{nonl}$ of the nonlinear terms in the interpretive rules. It can be seen that among all the eight nonlinear terms, seven are on the increasing tendency along with the training. The most drastic changes are $\sin(\delta_G)$ and $\cos(\delta_G)$ extracted from the transformation from *d-q* coordinate to natural coordinate, which have increased by about 33 (from 0.02 to 0.65) and 34 (from 0.02 to 0.68) times. The changes on the triangularly transforming of the generator power angle is easy to understand. In this research, we focus on the power angle stability, so the most direct nonlinear transformations about the power angle ought to be important factors for transient stability. However, at the beginning of the training, the generated interpretive rules are unreliable since the neural network evaluation model cannot predict the stability probability accurately and doesn't capture the key nonlinear terms. Along with the training, despite the accuracy of the hard labels (stable or unstable) predicted by the neural network evaluation model has increased from 74.3% (epoch=1) to 92.4% (epoch=10), the predicted probability has just surpassed the level of 50%. According to Fig. 9, although there has been an improvement on the frequency of $\sin(\delta_G)$ and $\cos(\delta_G)$, it is still relatively low. When the training epoch reaches 20, the accuracy of the well-trained neural network evaluation model is 0.943 and the hidden pattern has been captured accurately. As a result, the frequency of $\sin(\delta_G)$ and $\cos(\delta_G)$ is relatively high as shown in Fig. 9. In the whole, the average frequency of all the 8 nonlinear terms increased from 10.15 to 13.16 and only $V_B^2 P_B$ decreased, which means the voltage quadratic transform of bus active load is not an important nonlinear transformation under the TSA-ENRT framework, and excessive attention to $V_B^2 P_B$ may lead to even worse performance.

Besides, the information content of the nonlinear terms in the tree model has also been analyzed. According to [31], whether it is decision tree or regression tree, the splitting of nodes is performed in a sequential manner, and the order reflects the information content of the nodes. Different from Gini coefficient used in decision tree, the splitting criteria for a regression tree is mean squared error (MSE). However, the difference of MSE between datasets is large, hence, the average layer number $\bar{l}_{nonl}$ of the nonlinear terms is used to quantify the information content here for generalization, as shown in Fig. 10. It can be seen that the average layer numbers of $\sin(\delta_G)$, $\cos(\delta_G)$ and $V_B^2 Q_B$ have decreased. The decrease in the average layer number $\bar{l}_{nonl}$ indicates the importance of these three nonlinear interactions is increasing for the tree model. Compared with Fig. 9, sin($\delta_G$), cos($\delta_G$) and $V_B^2 Q_B$ are also the top three with the fastest increasing on the average frequency. However, the average layer numbers for other five nonlinear interactions have increased, and the information content is gradually decreasing, which seems contradictory to Fig 9. In fact, the average layer number and frequency are two different properties of the nonlinear terms in the generated interpretive rules, and the former represents information content, while the latter is related to the accuracy of regression. Some of the nonlinear terms, such as $\left(P_B \cos\theta_B + Q_B \sin\theta_B\right)/V_B$, have high frequency to ensure the regression precision, but they tend to appear at the end of the interpretive rules and only provide a small amount of information. Combining Fig. 9 and Fig. 10, it can be seen that the most important nonlinear terms are sin($\delta_G$), cos($\delta_G$) and $V_B^2 Q_B$. The other five nonlinear terms, although with low information content, work together to ensure the accuracy of the regression.

Finally, we have also analyzed the influence of different nonlinear interaction combinations on TSA-ENRT performance with $\sigma_{tree} = 10^{-2}$, as shown in Table 7. The nonlinear terms extracted from the transformation from *d-q* coordinate to natural coordinate, load power equation and load characteristic equation are abbreviated as *Non*$_1$, *Non*$_2$ and *Non*$_3$. It can be seen that the combination of the nonlinear terms has little influence on the accuracy of the neural network evaluation model, and mainly affects the performance of the tree model. The fidelity of TSA-ENRT is 0.016 larger than not using expert guiding nonlinear interactions. By comparing the results of rows 5-8, it can be seen that *Non*$_2$ has the greatest contribution, while the contributions of *Non*$_1$ and *Non*$_3$ are similar. The experiments of pairwise combination of nonlinear terms have also verified the above viewpoint. It's worth mentioning that the situation without nonlinear interactions does not mean that it will degenerate to GRU-TR. Due to the existence of the improved surrogate model training algorithm proposed in Section 3.3, even if we use the linear regression tree, the performance on consistency is better than GRU-TR by improving the fidelity for about 0.02 under the same tree regularization strength. The improvement also proves the rationality and effectiveness of the sample selecting and data augmentation strategy during training the surrogate model.

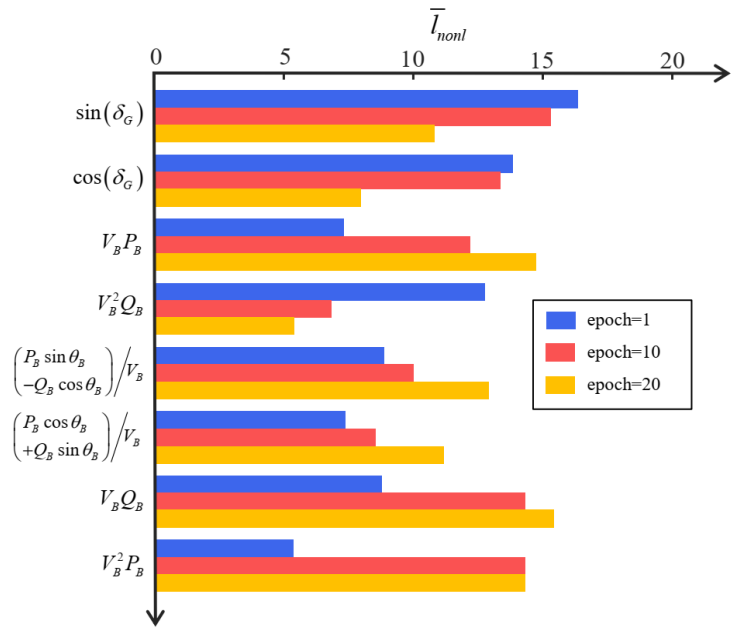

Fig. 10 The average layer number of the nonlinear interactions.

Table 7 Performance of TSA-ENRT with different nonlinear terms combinations.

| $Non_1$ | $Non_2$ | $Non_3$ | Acc | | Fidelity |
|---|---|---|---|---|---|
| | | | TSA-ENRT | TSA-ENRT-RT | |
| ✓ | ✓ | ✓ | 0.971 | 0.832 | 0.847 |
| × | ✓ | ✓ | 0.971 | 0.831 | 0.846 |
| ✓ | × | ✓ | 0.970 | 0.829 | 0.842 |
| ✓ | ✓ | × | 0.970 | 0.831 | 0.846 |
| × | × | ✓ | 0.970 | 0.828 | 0.842 |
| × | ✓ | × | 0.971 | 0.831 | 0.845 |
| ✓ | × | × | 0.971 | 0.828 | 0.841 |
| × | × | × | 0.970 | 0.818 | 0.831 |

## 4.6 Interpretive Rule Visualization

Fig. 11-13 are the generated interpretive rules corresponding to the predicted stability

probability of a sample whose ground truth is stable in different training stages. In the early stages of training, the sample was misclassified as unstable with 49.9% predicted stability probability since the poor performance of the neural network evaluation model and the corresponding interpretive rule is as shown in Fig. 11. Along with the training, the mistake was corrected at the $14^{th}$ epoch and the stability probability became 58.7%. Fig. 12 depicts the corrected interpretive rules. From the two interpretive rules, it can be seen that despite of the small change on the stability probability, the change of predicted stability can greatly influence the interpretive rule.

Besides, Fig. 12 and Fig. 13 present the generated interpretive rules from the same sample predicted as stable with different probability. In Fig. 12, the misclassification in Fig. 11 has been corrected with 58.7% stability probability predicted by the neural network evaluation model. Apparently, the judgment of the neural network evaluation model is still ambiguous. Along with the training, the predicted stable probability of the sample becomes 96.0%, and the corresponding interpretive rule is depicted as Fig. 13. It is easy to find that the two interpretive rules are not exactly the same, although both of these situations have been predicted as stable. Based on this, we can infer that even for the same category, different predicted probability corresponds to different interpretive rules, and our design of regression tree is reasonable.

What's more, the nonlinear terms extracted from the expert knowledge base have participated in the generation of the interpretive rules, which have been marked in blue. These nonlinear terms with explicit physical significance can better explain the nonlinear process of the neural network. The physical significance makes the interpretive rules more fit for human cognition and improves the interpretability.

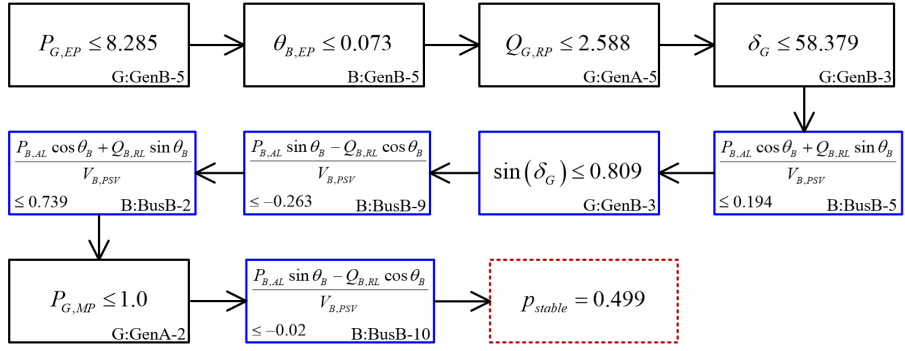

Fig. 11 Interpretive rule of a stable sample with 49.9% predicted stability probability.

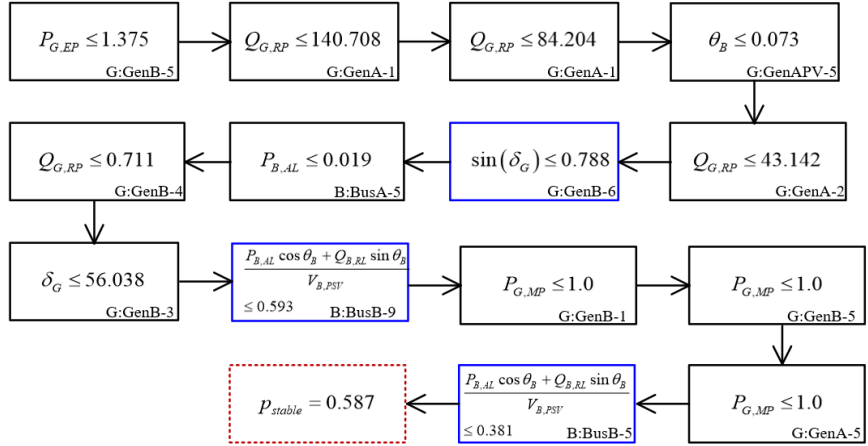

Fig. 12 Interpretive rule of a stable sample with 58.7% predicted stability probability.

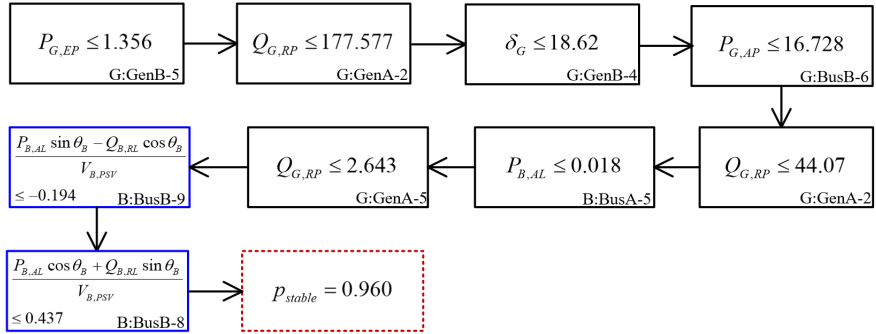

Fig. 13 Interpretive rule of a stable sample with 96.0% predicted stability probability.

# 5 Conclusion

In this paper, an interpretable power system transient stability assessment method with expert guiding neural-regression-tree (TSA-ENRT) has been proposed, which combines expert knowledge base, neural network evaluation model, nonlinear regression tree and tree regularization surrogate model. The nonlinear regression tree is generated by extracting expert knowledge from a simple two-machine three-bus system to approximate the stability probability predicted by the neural network evaluation model. The interpretive rules generated by the expert guiding tree model have human-understandable nonlinear terms with clear physical meaning and are closer to the nonlinear nature of neural networks compared with linear decision tree. By approximating the prediction of the neural network evaluation model in a regression way, the interpretive rules can establish a connection between the probability information from the network evaluation model and the tree model. By using the tree regularization, the accuracy of the neural network evaluation model and the interpretability of the tree model can be balanced with mutual trade-off, and the introduction of expert knowledge based nonlinear terms can generate human comprehensible rules. Extensive experiments on the testing power system demonstrate the superior performance of TSA-ENRT on accuracy, fidelity and interpretability.

## Acknowledgement

This work is supported by The National Key R&D Program of China "Response-driven intelligent enhanced analysis and control for bulk power system stability" (2021YFB2400800).